# Ultrasensitive and fast single wavelength plasmonic hydrogen sensing with anisotropic nanostructured Pd films

William L. Watkins[a,*] and Yves Borensztein[a,*]

[a] Sorbonne Université, CNRS, Institut des NanoSciences de Paris, INSP, F-75005, Paris, France

Anisotropic nanostructured porous Pd films are fabricated using oblique angle deposition in vacuum on a glass substrate. They display a dichroic response, due to localised surface plasmon resonances (LSPR) within the nanoparticles forming the film, dependent on the incident light polarisation. Ultrasensitive hydrogen sensing is reached by using these films in conjunction with a differential optical technique derived from the reflectance anisotropy spectroscopy. The evolution of the samples' optical responses is monitored during the formation of Pd hydride in both the dilute α-phase and the dense β-phase, whilst the samples are exposed to different concentration of $H_2$ in Ar (from 100% $H_2$ to a few ppm). The measurements are performed at a single wavelength in the visible range and at 22°C. The results show that a quantitative measurement of the hydrogen concentration in a carrier gas can be measured throughout the concentration range. The limit of detection is 10 ppm and the time for detecting the presence of $H_2$ in the carrying gas is below one second at concentration down to 0.25% of $H_2$ in Ar. Furthermore, the optical anisotropy of the samples and its evolution with exposure to $H_2$ are correctly reproduced with an effective medium theory.

Corresponding Authors: *william.watkins@insp.jussieu.fr, *yves.borensztein@insp.jussieu.fr

## 1. Introduction

The development of hydrogen technology is today one of the main path to the production of clean and renewable energy [1–4]. However, the use of dihydrogen rises concerns from its explosive properties in air. 4% of $H_2$ is indeed the flammable mixture and thus efforts are made to develop selective, sensitive and fast responding dihydrogen detectors [5,6]. Several paths are being investigated to reach this goal [7–11] in particular with sensors based on Pd, as it selectively reduces when exposed to $H_2$ to form palladium hydrides [12]. In recent years, attempts have been made to develop plasmonic sensors, using localised surface plasmon resonances (LSPR) excited by light in metal-Pd or pure Pd nanoparticles (NPs), based on the change in optical response of Pd when forming hydrides [13]. In addition to the good sensitivity and fast response of such plasmonic sensors, the use of light as a probe implies that the sample is a passive component of the system, which eliminates hazards such as electrical sparks associated with electronic setups.

Most LSPR-based sensors [14–20] are made of Au NPs which exhibit a well-defined plasmonic resonance. Au NPs having weak interaction with $H_2$ [21], hybrid systems such as core-shell Au-Pd NPs [22,23], AuPd alloy NPs [20], Au NPs - Pd NPs oligomers [24,25] have been used. In these systems, the hydrogenation of Pd modifies the dielectric environment of the Au NPs, hence induces a shift in wavelength of the LSPR of Au, which gives a measurement of the presence of $H_2$ [13]. Recently, K. Sugawa et al. have shown Pd to be the "third plasmonic sensing material". Indeed, LSPR in the visible range has been obtained by using large Pd spheres (diameters of 100 to 200 nm), displaying a dependency to the embedding medium's refractive index change, demonstrating the ability of Pd particles for plasmonic sensing [26]. It has also been shown that flat Pd nanodisks and nanorings of sizes around 300 nm, prepared by colloidal lithography, display LSPR in the near infra-red, which is red-shifted upon exposure to $H_2$, due to the formation of Pd hydride [27].

Despite this sensitivity to $H_2$, the thermodynamics of Pd hydride does impose an important limitation for hydrogen sensing applications. Indeed, the phase diagram of Pd hydride exhibits different phases depending on the partial pressure of $H_2$, $p(H_2)$. At room temperature, in the case of bulk Pd, the dense phase, $PdH_x$ with $x \geq 0.6$, known as the β-phase, is obtained for $p(H_2)$ larger than about $10^{-2}$ bar, i.e. more than about 1% $H_2$ in ambient gas at atmospheric pressure. On the other hand, at $p(H_2) \lesssim 10^{-3}$ bar, only the low density α-phase is reached, with a maximum value $x \approx 0.02$ [28]. At intermediate pressure, a hybrid α + β phase is formed, constituted by β-phase regions embedded in the α-phase [29]. Regarding sensing applications, concentrations much lower than the flammability limit of 4% have to be detected, and it is therefore mandatory that $H_2$ sensors be sensitive to the very initial α-phase formation [13]. This leads to an important issue for plasmonic sensors, as only the optical response of the dense β-phase strongly differs from that of pure Pd [30], whereas the optical response of the α-phase is very close to Pd. Consequently, shifts in the LSPR position large enough to be easily observed with conventional plasmonic sensors are reached only when the β-phase or the α + β phase are obtained, i.e. for $p(H_2)$ larger than about $10^{-3}$ to $10^{-2}$ bar at room temperature.

In the range of interest $p(H_2) \leq 10^{-3}$ bar, the α-phase leads to sub-nanometre shifts of the LSPR, which are difficult to measure with conventional plasmonic sensors. Indeed, the shift is usually determined by performing spectroscopic measurements around the resonance, followed by a fitting of the obtained spectra. Clearly, such a method is limited by the resolution and the reproducibility of the spectrometers which are, for current monochromators, depending on the width of the monochromator's slits, between 0.1 nm and a few nm. Improved resolution as low as 0.01 nm, necessary for detecting the α-phase, could be reached by using high-resolution monochromators with micrometre slit, though these apparatus are bulky and expensive, thus not suitable for simple, low-cost and easy-to-use sensors.

In order to answer these issues and to increase the sensitivity of plasmonic sensors, several authors developed specific Au/Pd or Pd nanostructures**.** For example, studies conducted by Langhammer's group showed a good sensitivity and reproducibility at 2% and 4% of $H_2$ in Ar as carrying gas for Pd NPs and Au-Pd heterodimers placed in a flow reactor [31,24]. Yang et al. reported a detection limit down to 2% $H_2$ in $N_2$ by using Au-Pd dimers and trimers [25]. With a "Pd-based plasmonic perfect absorber", Tittl et al. were able to detect concentration down to 0.5% [32]. Jiang et al obtained a detection limit as low as 0.2% for $H_2$ in $N_2$ with bimetallic Au/Pd nanorods [23]. Langhammer et al. [33] and Wadell et al. [20], indeed presented results with detection limit down to 0.1 % of $H_2$, but the experiments were performed in a vacuum chamber and not at atmospheric pressure. To our knowledge, no LSPR-based system showed detection limit for partial pressures of $H_2$ in a carrying gas far below 1%, i.e. in the range of early formation of the α-phase, at or below 0.1%. Such sensitivity is desirable, for safety consideration but also for analysis of impurities in several industrial processes [6], as it has been brought forward by Wadell et al. [13].

In the present investigation, we address the sensitivity issue due to the resolution limitation of the monochromator, by using the Transmittance Anisotropy Spectroscopy (TAS), derived from the Reflectance Anisotropy Spectroscopy, which has been shown to be a very efficient and sensitive optical technique for investigating pristine or adsorbate-covered crystalline surfaces [34–38] and for supported metal NPs [21,39–41]. This method being differential, any kind of instability is eliminated, should it be due to the fluctuation in the light source or to mechanical noise. There is also no interference with ambient light and it can be used in illuminated environment. This technique leads to very stable and very sensitive measurements, which are not reachable with conventional plasmonic methods. Additionally, this apparatus is operated at a single wavelength and does not require spectroscopic measurements, which therefore frees oneself from the use of a monochromator and strongly lightens the operating system. Such an approach has been previously followed with anisotropic Au NPs dimers for the detection of biomolecules [42]. Similarly, by use of polarisation-dependent Au-Pd heterodimers, Wadell et al. have increased their sensitivity to $H_2$ detection and suppress drift issues [24]. However, in these examples, the NPs were prepared by lithographic methods, and the large size of the NPs reduced the overall detection limit to $H_2$ exposure. It has been shown indeed that the smaller the nanoparticles, the greater the surface-to-volume ratio, and the higher the sensitivity and the shorter the response time [13,43,44].

Consequently, efforts are to be made to develop smaller but anisotropic NPs, in order to exhibit fast response to hydrogen and be investigated by the TAS technique. We address this second issue by elaborating pure Pd films, grown by oblique angle deposition, as already proposed for other metals [45,46]. This method provides anisotropic nanostructured films formed by NPs a few nm large, where the LSPR excited by the impinging light strongly depends on the polarisation of light, leading to dichroic properties suitable for TAS. We demonstrate that this yields a sensitivity to amounts of $H_2$ diluted in Ar as small as a few ppm at room temperature, and also a fast response in the range of seconds. Moreover, it also grants us the ability to determine the precise amount of small fractions of $H_2$ diluted in Pd, showing the quantitative potentiality of this technique. Experiments were also undertaken in dry and humid air (50% humidity) to determine perspectives in terms of selectivity and reactivity of the system in realistic conditions.

## 2. Experimental details

The elaboration of the anisotropic Pd films is done using oblique angle deposition under a $3.10^{-6}$ mbar vacuum on glass substrates, prepared from microscope slides cleaned with ethanol. The evaporation is carried out from a crucible heated by direct current with the samples positioned at a grazing angle of about 10° to the crucible and the evaporation rate was 0.2 nm/s. The mass thickness is controlled using a calibrated quartz balance and corrected to account for the evaporation angle. The characterisations of the samples are carried out by scanning electron microscopy (SEM) and optical measurements. The SEM apparatus used is a Zeiss Supra 40. The settings are EHT = 5 kV at a working distance of 2.8 mm with an aperture of 7 µm. The SEM needs to be adjusted with a low potential, as the substrate being nonconductive, the samples are prone to charging. The optical absorbance was measured on a Agilent Cary 5000 UV-Vis-NIR spectrometer. Both parallel and perpendicular polarisations in relation to the sample's direction of evaporation are recorded. The anisotropic measurements are performed on a homemade TAS system. The TAS technique used here is derived from the RAS apparatus with Aspnes configuration [47], and measures the transmission anisotropy of transparent samples. Further details of the experimental setup can be found in Ref. [21]. The samples are investigated in a gas flow reactor equipped with a silica window enabling the TAS measurements during gas cycles. After purging of the cell with pure Ar for 24 hours in order to eliminate any possible contamination, the samples are exposed alternately to pure Ar and to various $H_2$/Ar mixtures at atmospheric pressure. For this purpose, different concentrations of $H_2$ in Ar are prepared in a dedicated bottle by the following procedure. The mixture bottle is vacuumed whilst heated to $3.10^{-3}$ mbar for 24 h. It is then filled with 2.5 bar of $H_2$ and completed to 4 bar with Ar. The bottle is then emptied to atmospheric pressure and refilled to 4 bar with Ar. By using this procedure several times, accurate proportions of $H_2$ in Ar down to a few ppm can be obtained. The experiments in the gas flow reactor are undertaken at atmospheric pressures and at 22°C, and the gas flows are fixed to 1000 sccm. These large flows are used to rapidly switch from pure Ar to the chosen concentration, and vice-versa. For measurements performed with dry or with humid air as carrying gas, a similar procedure for preparing the mixture is used.

## 3. Results and discussion

### 3.1. Microscopic and optical characterisations

The SEM image of a typical Pd film, obtained by vacuum deposition of Pd with mass thickness approximately equal to 1.9 nm is shown in Fig.1.a. The nanostructured Pd film is porous, and made up of elongated islands, separated by trenches. These islands appear to be formed by agglomerated NPs of size around 10 nm, separated by narrow gaps, and they are in majority oriented close to the direction normal to the evaporation orientation, indicated by the arrow. The morphological anisotropy of the film is better seen in the 2D Fourier transform of the SEM image (Fig.1.b).

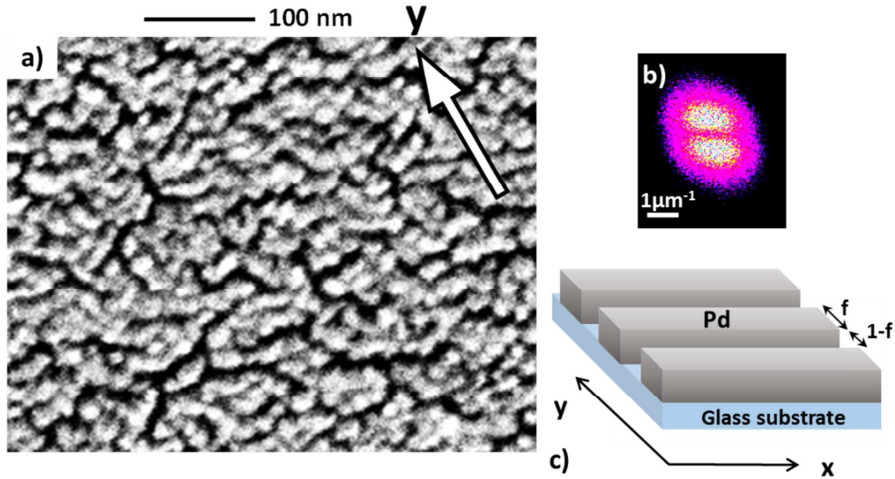

**Fig.1.** (a): SEM image of the porous Pd film. (b): 2D Fourier transform of the SEM image. (c): scheme of the Pd blocks and narrow gaps used for modelling the optical response of the film. The arrow indicates the orientation of the vacuum deposition.

The optical properties of the film are shown in Fig.2, where the absorbance, given by $-log(T)$, where T is the transmission of the sample, is drawn for light polarisation either parallel or perpendicular to the direction of evaporation. It shows a clear dichroism. The spectrum measured under parallel polarisation displays a broad resonance centred around 500 nm, whereas the spectrum for perpendicular polarisation does not exhibit such a maximum. The maximum of absorbance around 500 nm is due to LSPR located within the Pd film. Plasmonic absorptions have been previously observed in porous films and in films formed by agglomerated NPs for other metals, e.g. Ag, Au or Al, and have been ascribed to LSPR located on the nanoparticles in close interaction or/and on the voids within porous and granular films [48–54]. Accurately modelling the optical response of such an irregular film on a substrate is a huge challenge, and would at least demand the exact knowledge of the shape, size and distribution of the interacting NPs in the islands, and also to account for multipolar effects and interaction with the substrate [50,51,54–58]. Consequently, a simplified theoretical model is used in this study to reproduce the experimental data. The islands forming the nanostructured porous Pd film are represented by distributions of elongated metal blocks separated by gaps (as indicated in the schematic of Fig.1.c). The optical response and the LSPR supported by such a lamellar film, can be obtained within an effective medium theory, as described by Aspnes [59]. Two effective dielectric functions are then obtained describing the film's optical response in the two main directions, x and y, of the slab:

$$\varepsilon_x(\omega) = f\varepsilon_{Pd}(\omega) + (1-f)\varepsilon_m \qquad (1.a)$$

$$\varepsilon_y(\omega)^{-1} = f\varepsilon_{Pd}^{-1}(\omega) + (1-f)\varepsilon_m^{-1} \qquad (1.b)$$

where $\varepsilon_{Pd}(\omega)$ is the dielectric function of bulk Pd and $\varepsilon_m$ the dielectric function of the empty gaps, i.e. 1 for gas. $0 < f < 1$ is the fraction of Pd in the film.

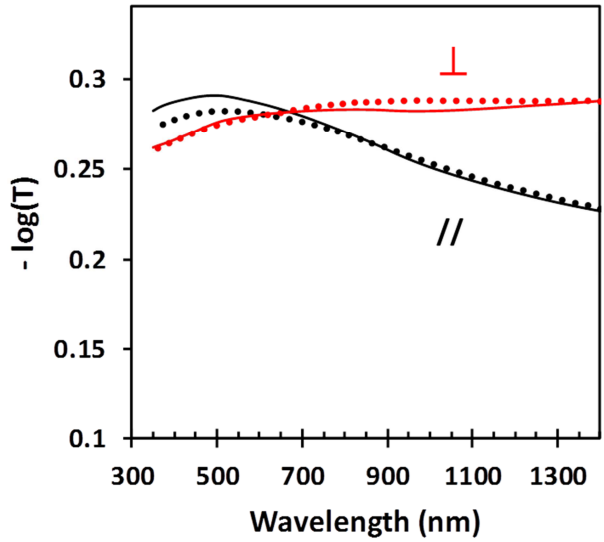

**Fig 2.** Dotted curves: experimental absorbance (-log (T)), for polarisation parallel (black) and perpendicular (red) to the direction of evaporation. Continuous curves: result of calculation.

No resonance (related to a maximum in $\varepsilon_{x,y}(\omega)$) can be obtained for x polarisation, whereas, for y polarisation, LSPR localised in the gaps are obtained for values of $\varepsilon_{Pd}(\omega)$ given by the pole of $\varepsilon_y(\omega)$, i.e. by: $\varepsilon_{Pd}(\omega) = -f/(1-f)$. The dielectric function of Pd has a shape similar to that of a Drude function and the LSPR is located in the visible/infrared range for values of f ≥ 0.9 (See the Supplementary Material). Typically, the LSPR is at 350 nm for f = 0.9 and at 500 nm for f = 0.95. However, the calculation for a given f leads to a narrower and sharper resonance than what is observed (Fig.S.2). In order to reproduce the observed broadening of the LSPR, the actual disordered nanostructure of the sample is taken into account by considering a distribution of values for f. Using the dielectric function experimentally obtained for Pd by Johnson and Christy [60] allows one to nicely reproduce the experimental absorbance, as shown in Fig.2 where the calculated absorbances in both directions are drawn in continuous lines. The mass thickness of the film is also obtained and found to be 1.4 nm, close to the value estimated from the quartz balance. The maximum of absorbance around 500 nm corresponds to polarisation of light normal to the gaps, with a distribution of f ranging from 0.85 to 1. It is worth noticing that the Pd film could have been modelled in a different simplified way: the islands formed by the small agglomerated NPs could be considered as being larger elongated NPs, which also display shape-dependent LSPR. Details of the calculations and deeper discussion are given in the Supplementary Material.

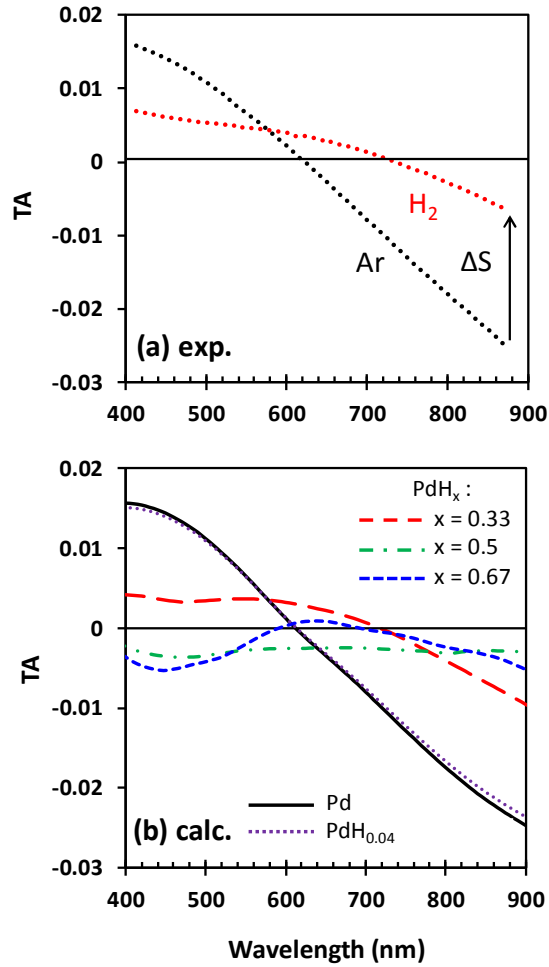

Fig 3. (a) Transmission anisotropy of the Pd film measured in pure Ar (black dots) and in pure $H_2$ (red dots). (b) Calculated spectra for pure Pd (continuous black curve), $PdH_{0.33}$ (long-dashed red curve), $PdH_{0.5}$ (short-dashed green curve) and $PdH_{0.67}$ (short dashed blue curve); the purple dotted curve is the calculated spectra for $PdH_{0.04}$.

### 3.2. Exposure to atmospheric pressure of Ar and H2

By using the TAS method, the transmission anisotropy (TA), $\Delta T/T$, is directly measured on the sample placed in the gas flow reactor and is given by:

$$\frac{\Delta T}{T} = \frac{T_\perp - T_\parallel}{1/2(T_\perp + T_\parallel)} \qquad (2)$$

where $T_\perp$ and $T_\parallel$ are the transmissions of the sample for polarisations perpendicular and parallel to the evaporation direction. The TA spectrum measured under pure Ar at atmospheric pressure is drawn in Fig.3.a. It confirms the optical anisotropy shown in Fig.2, for absorbance measurements in both polarisations. The TA goes through zero as the two spectra of Fig.2 cross each other. When the sample is exposed to pure $H_2$ at atmospheric pressure, the TA spectrum also drawn in Fig.3.a. is strongly modified and its intensity is reduced.

After purging the reactor with Ar, the signal reverts back to its initial state, hence showing the perfect reversibility of the phenomenon. The effect of $H_2$ exposure can be ascribed to the absorption of hydrogen by the Pd NPs, leading to the formation of the β-phase $PdH_x$ and, consequently, to a strong change in its dielectric function. In order to reproduce the TA spectrum under $H_2$, it is necessary to use the dielectric function of the β-phase $PdH_x$. However, to our knowledge, no reliable dielectric functions of hydrogenated Pd have been determined experimentally. Instead, using first principle methods, Silkin et al. have calculated the dielectric functions for pure Pd and hydrogenated $PdH_x$ with different values of x [30]. The important aspect here is to show that the change in the dielectric function of the nanostructured Pd film, due to formation of hydride, does explain the experimental observation. The TA spectrum measured under Ar (Fig.3.a) is reproduced with a Pd anisotropic nanostructured film and Silkin's dielectric functions following analogous steps as shown in the previous paragraph. The result given in Fig.3.b is in good agreement with the experimental TA of Fig.3.a. The obtained parameters describing the nanostructured anisotropic film are very similar, although not identical, to those previously obtained with the experimental Johnson and Christy dielectric function [60], as explained in the Supplementary Material. Keeping these parameters and replacing the dielectric function of Pd with those of $PdH_{0.33}$, $PdH_{0.5}$ and $PdH_{0.67}$ leads to a strong decrease of the TA, as shown in Fig.3.b. A very good agreement is found between the experimental TA spectrum under $H_2$ and the calculation for $PdH_{0.33}$, though the expected Pd hydride should be closer to $PdH_{0.5}$ or $PdH_{0.67}$. This result can be due to the difficulty in accurately determining the dielectric functions of crystals with ab-initio methods. Indeed, discrepancies exist between the experimental and the calculated dielectric functions for pure Pd, respectfully obtained by Johnson and Christy and by Silkin et al. (see the Supplementary Material). It can also be wondered whether the known expansion of $PdH_x$ with respect to pure Pd could explain this difference. [29] However, we verified that taking into account this effect in the calculation did not strongly modify the results for $PdH_{0.5}$ or $PdH_{0.67}$. Finally, it is also likely that the actual H concentration is smaller than in bulk: it has indeed been shown that H concentration within Pd at a given $H_2$ pressure is smaller for NPs than for bulk [61]. Nevertheless, the above comparison does show that the observed change of the TA signal is indeed due to the formation of Pd hydride, which modifies and reduces the LSPR in the Pd gaps. This illustrates that the present nanostructured porous films are quite reactive to $H_2$ absorption and can be used for sensing smaller amounts of $H_2$, even when solely the α-phase is obtained, as shown in next paragraph.

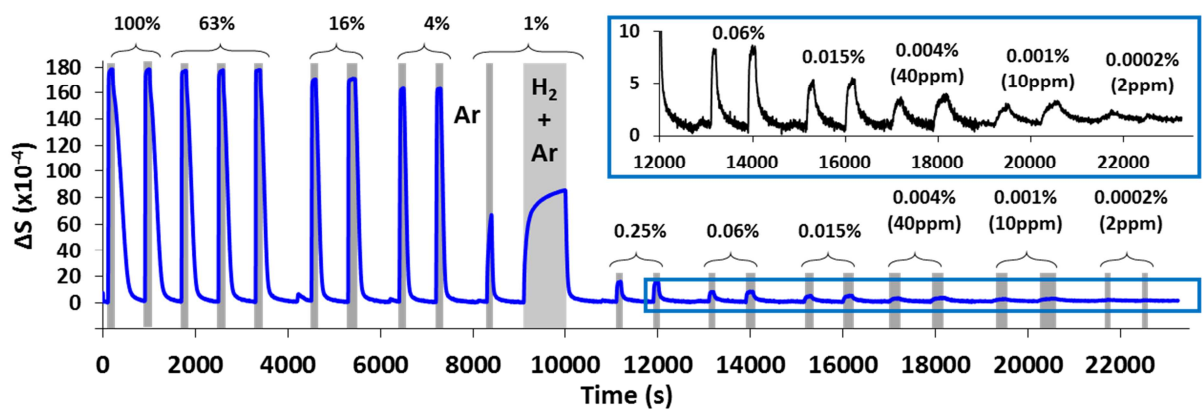

**Fig 4**. Recording of the TA signal measured at the wavelength 885 nm, during alternative cycles between pure Ar and different decreasing concentrations of $H_2$ in Ar, indicated on the graph. The signal close to zero is measured in pure Ar (white areas in the graph), while the positive signal is obtained under $H_2$/Ar mixtures (grey areas). The insert gives a zoom for small concentrations.

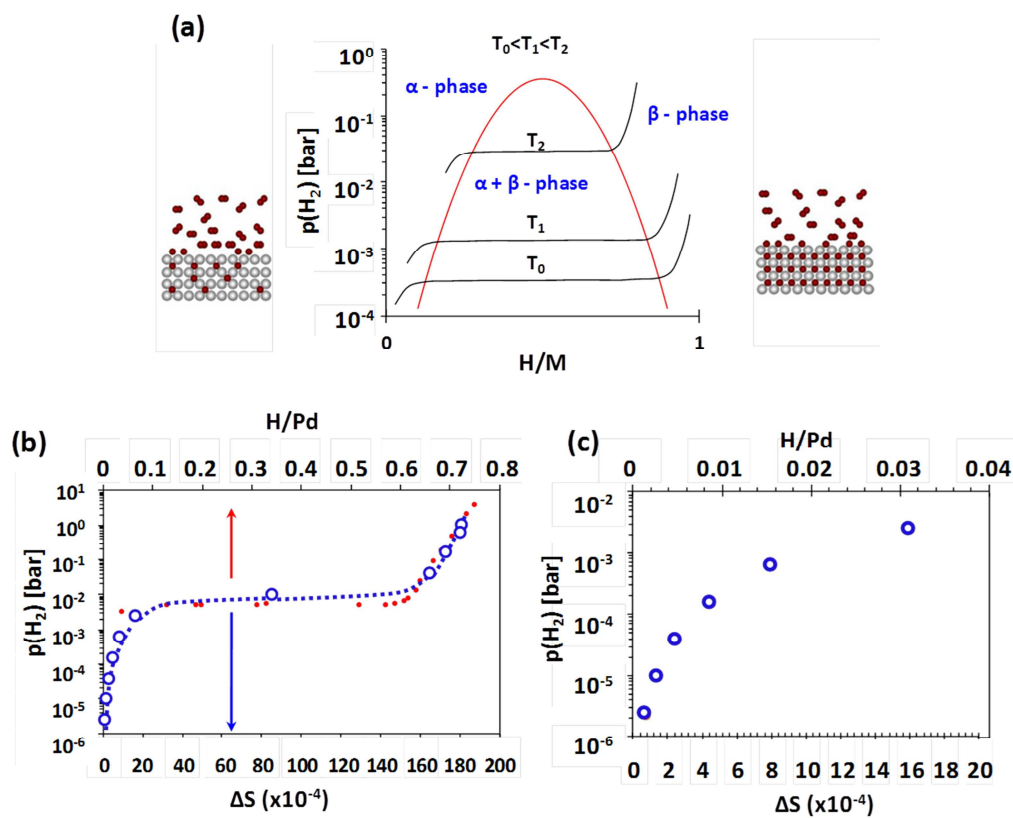

**Fig 5**. (a) Scheme of the phase diagram showing the α-phase, the β-phase and the plateau where both phases coexist. (b) Graph showing the partial pressure of $H_2$ in Ar as a function of the amplitude of the signal measured in Fig 4. The phase diagram obtained at 22°C for bulk Pd is drawn for comparison (from Ref. [28]). (c) : zoom of (b) for the smaller values of $p(H_2)$, with the x values determined from calculation, in log-log representation.

### 3.3. Pressure cycles of Ar and $H_2$ in Ar

The real-time experiments consists in measuring the change of the TA signal, $\Delta S$, as a function of time, at a given wavelength, during exposure cycles to pure Ar and $H_2$ in Ar with decreasing concentrations. The chosen working wavelength is where the largest change in TA within the investigated wavelength range is observed, therefore $\lambda$ = 885 nm (see Fig.3.a). An optical offset is added in order to have a TA value close to zero under Ar to increase the sensitivity of the measurements. The experiments are performed with reducing partial pressures from 1 bar to $2 \cdot 10^{-6}$ bar of $H_2$ in Ar at atmospheric pressure, which corresponds to $H_2$ concentration in Ar varying from 100% to 0.0002% (2 ppm). The results are drawn in Fig.4. Similar results are obtained with other Pd samples.

When the gas is switched from Ar to $H_2$ in Ar, the signal change $\Delta S$ is almost immediate and positive, as expected from Fig.3.a. When exposed back to Ar, the signal reverts quickly, and eventually the sample regenerates to its initial state. Fig.4 shows that the sample follows three different behaviours. At high concentrations of $H_2$ in Ar, between 100% to 4%, $\Delta S$ is large and quickly saturates within a few seconds or less; its intensity is almost unchanged throughout the range. At low concentrations, 0.2% and below, the signal is about 10 to 100 times smaller. The intensity depends on the concentration of $H_2$, but saturation is also rather quickly reached. Fig 4 also shows that the detection limit in the present case is about 2 to 10 ppm. At 1% though, the signal behaves differently; the increase is also fast, but saturation is not reached, even after 1000 s under $H_2$ in Ar. The intensity is also intermediate between the high and the low pressure regimes. These three behaviours correspond to the different Pd hydride phases that are schematised in Fig.5.a for different temperatures, where the relation between the pressure and the proportion x of H in PdH$_x$ is drawn. The region on the left hand side of Fig.4 corresponding to $H_2$ percentages between 100% to 4% (1 to 0.04 bar) correlates to the formation of the dense β-phase. The region on the right hand side, corresponding to percentages from 0.2% (0.002 bar) and lower, correlates to the formation of the dilute α-phase. The experimental point at 1% corresponds to the co-existence of both phases, where the β-phase is progressively formed.

The $H_2$ partial pressures as a function of the maximum signal change $\Delta S$ reached for every pressure is displayed in Fig.5.b. It shows that the shape is perfectly analogous to the phase diagram in Fig.5.a. The phase diagram for bulk Pd determined at 20°C is also drawn as a function of the H/Pd ratio x in the hydride (upper abscise) on Fig.5.b (taken from Ref. [29]). The scale of this latter axis has been chosen so the phase diagram obtained in our experiment (blue circles) is superimposed with the bulk phase diagram (red dots). This demonstrates that the observed change in the TAS signal can provide a measurement of the H concentration in the Pd film after calibration, although it cannot be inferred from this superposition that the H/Pd ratio x indicated on the upper abscise does exactly correspond to our experimental results. It has indeed been shown that, for nanoparticles, the phase diagram is very similar, although the shape and the width of the terrace, and the H concentration, depend on their size [33,61] .

A way to even more enhance the sensitivity of such a system and to determine amounts of the order of 1 ppm or less, could be to increase the optical anisotropy of the Pd film. This is shown in the Supplementary Material (paragraph S.4), where the TAS measured for three different Pd samples are shown, together with the cycles obtained when switching from Ar to 4% $H_2$ in Ar. The anisotropy is related to the morphology of the film, which in turn is related to the preparation conditions (thickness of the Pd film, angle of the deposition, deposition rate, deposition temperature, possible annealing temperature...). A complete investigation of these parameters would consequently be important for the development and optimisation of this method but is out of the scope of the present article.

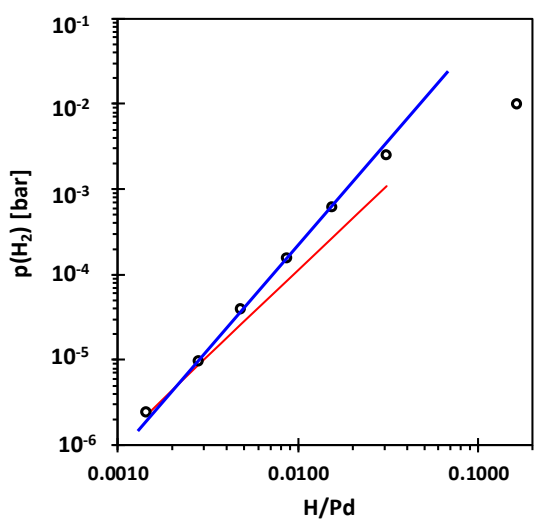

**Fig 6.** Plot of the pressure of $H_2$ against the H/Pd ratio in log-log scales. The black dots are the experimental data. The blue line gives a fitting of the data with a slope of 2.36. The red curve shows Sieverts' law with a slope of 2.

In order to quantitatively relate ΔS to the H/Pd ratio x for low pressures of $H_2$, one must consider the dielectric function of dilute $PdH_x$ compounds calculated by Silkin et al. Although they did not present results for values smaller than x = 0.125, their dielectric functions for Pd and for $PdH_{0.125}$ being very close, it is reasonable to consider intermediate functions obtained by interpolation for x ≤ 0.125. The effect of a small value of x, e.g. equal to 0.04, two times larger than the density limit of the α-phase, is shown as an illustration in Fig.3.b. (purple dotted spectrum). The calculation procedure is explained in the Supplementary Material. This illustrates that the change of transmission signal for small amount of H in Pd is indeed weak, and that the present differential method based on the TAS is necessary to measure such minute changes. By using the above procedure, one can determine the values of x from the measurement of ΔS for every $H_2$ partial pressures in the α-phase. The so-obtained values are drawn as abscises in the upper part of Fig.5.c. for pressures lower than $2 \cdot 10^{-3}$ bar (0.2%). For $p(H_2) = 6 \cdot 10^{-4}$ bar, the H/Pd ratio obtained is 0.016. This almost corresponds to the complete α-phase. For the higher pressure of $p(H_2) = 2 \cdot 10^{-3}$ bar, the H/Pd ratio is 0.031. In this case, the α-phase has saturated and the β-phase has started to form. From that point on, this procedure for determining x is thus no longer valid as the change of the dielectric function must now account for the development of the β-phase.

The partial pressure of $H_2$ against the so-determined H/Pd ratio x, in the α-phase, is drawn in log-log diagram in Fig.6. It shows a linear relationship, with a slope equal to 2.36 up to x = 0.016. The deviation from the linear dependence, which starts slightly from the value of x = 0.031, confirms that the β-phase is starting to form. The point determined by the same procedure, for $p(H_2) = 10^{-2}$ bar, and which gives x ≈ 0.15, is clearly out of the linear evolution, which was indeed expected as, for this pressure, the two phases are coexisting. Coming back to the initial evolution, Sieverts' law [12] indicates that the H/Pd ratio x should be proportional to the square root of the pressure, i.e. the points should be aligned with a slope of 2 (slope = 1/0.5), which is drawn in Fig.6 (red). The measured slope (slope = 1/0.43) is thus in variance with the Sieverts' law. This deviation is most likely related to the complex dissociation process of $H_2$ molecules on the surface of the Pd NPs. While Sieverts' law corresponds to a second order, Langmuir-type, dissociation of $H_2$, i.e. that two empty sites are necessary for the reaction to take place, it has been shown that the dissociation on the (111) surface actually involves more than two empty sites and therefore that the reaction deviates from the expected Langmuir process [62,63]. However, a detailed analysis of this process is out of the scope of the present article.

**3.4. Response time**

As well as the detection limit, a second important aspect regarding $H_2$ sensing is its response time. Two ways can be considered to measure the response time. For qualitative measurements, which are crucial for safety consideration, one can define the time, $t_{detect}$, it takes for the sensor to provide a significant response, i.e. the time it takes to register the presence of $H_2$ in the atmosphere. On the other hand quantitative measurements conventionally consider the time, $t_{90}$, after which 90% of the signal is reached. According to European and US authorities, hydrogen sensors are required to have a response time of less than 1 s for industrial applications [64].

In Fig 7, the evolution of the TA signal measured as a function of time on another sample, is shown in detail for three concentrations of $H_2$: 63%, 4% and 0.25%. These are recorded at a faster rate of 0.1 s per point, and correspond in the two former cases to the explosive concentration range (and leading to the formation of the β-phase), and in the latter case to the sub-flammable range (and leading to the α-phase). For 63% $H_2$ in Ar, $t_{90}$ = 3.5 s, and $t_{detect}$ ≤ 0.2 s, whilst at 4% and 0.25%, $t_{90}$ increases to 32 s and 40 s and $t_{detect}$ to 0.6 s and 1 s, respectively. The measured response time $t_{90}$ is actually sample dependent, and varies, e.g. for 4% of $H_2$, from about 20 to 35 seconds. Table 1 summarises the values of $t_{90}$ and $t_{detect}$ determined for two samples. The obtained values for the response time $t_{90}$ at concentrations between 63% to 1% are better or of the same order as previously reported for Pd-based plasmonic sensing systems (10 s for 100% [25], 30 s for 4% [31], more than 60 s for 4% [24], 20 s for 100% and 60 s for 1% [65], less than 90 s for 0.2% [23]). Smaller response times ($t_{90}$ = 1 s) have been previously observed in plasmonic sensors using Pd, but not with $H_2$ in a carrying gas at atmospheric pressure; instead, the experiments were undertaken in vacuum, where the $H_2$ pressure was increased from vacuum to 40 mbar [20]. The increase of $t_{90}$ observed in the present experiments, from 3.5 s at 63% to 20 or 32 s at 4% is likely due to the competition between physisorption of $H_2$ molecules at the NP surfaces, prior their dissociation [66], and transient physisorption of Ar atoms. This competition between $H_2$ molecules and atoms or molecules of the carrying gas appears to be an intrinsic limitation which seems difficult to overcome for a faster quantitative determination of $H_2$ concentration in a gas. Results with air as a carrying gas, given in the next paragraph, reinforce this explanation. A second limitation for obtaining a fast response time occurs for partial pressures corresponding to the mixed α-phase / β-phase. This is quite clear from Fig.4, which shows that saturation is not reached at 1% of $H_2$, even after 1000 s. Nevertheless, the detection time is still below 1s at 1% and at 0.25%, i.e. below the flammability range and thus within the requirements for hydrogen sensors. For smaller partial pressures, $t_{detect}$ increases to about 10 s at 0.06% to 50 s at 10 ppm, which is still an acceptable detection time for security applications.

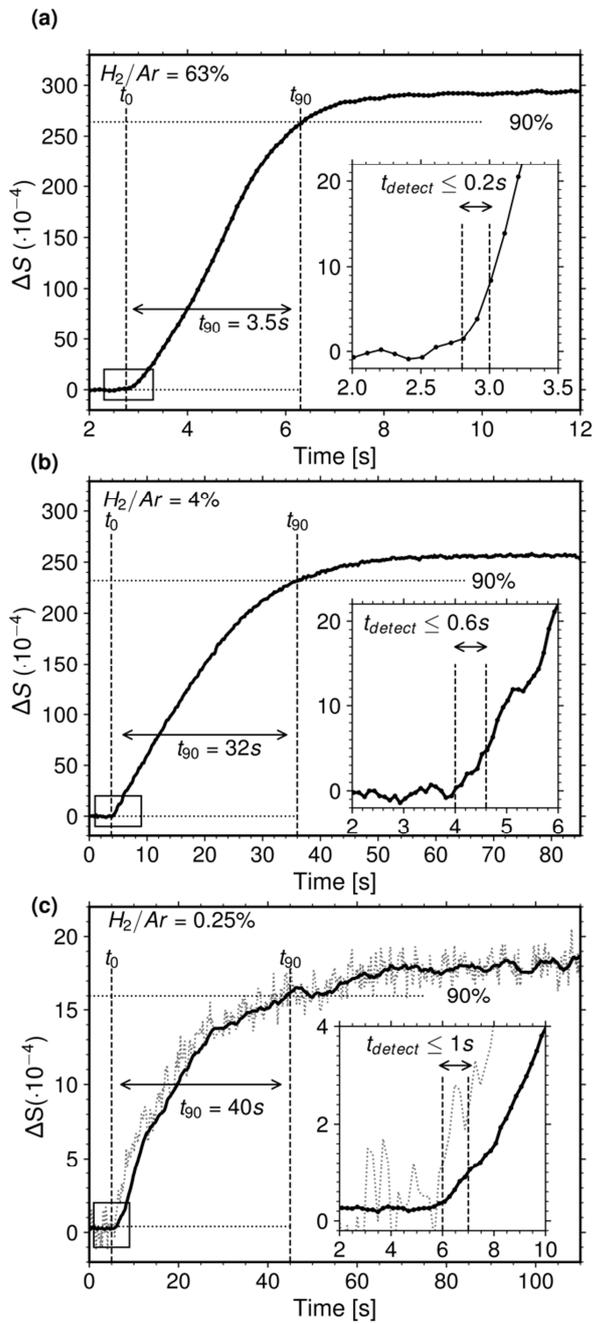

**Fig 7.** These three graphs show the response time of another Pd sample, recorded at a rate of 0.1 s per point, when exposed to (a) 63%, (b) 4%, and (c) 0.25% of $H_2$ in Ar. In this latter case, the grey line is at the same rate, the black line is a moving average of 2 s. The large graphs indicate $t_{90}$ whereas the inlets show the beginning of the response and thus $t_{detect}$.

| $H_2$ in Ar (%) | Sample 1 (recorded at a rate of 2s) | | Sample 2 (recorded at a rate of 0.1s) | |
|---|---|---|---|---|
| | $t_{detect}$ (s) | $t_{90}$ (s) | $t_{detect}$ (s) | $t_{90}$ (s) |
| 63 | < 2 | 4 | 0.2 | 3.5 |
| 16 | < 2 | 9 | 0.3 | 7.5 |
| 4 | < 2 | 20 | 0.6 | 32 |
| 1 | < 2 | - | 1 | - |
| 0.25 | < 5 | 30 | 1* | 40 |
| 0.06 | 7 | 35 | 10* | 70 |
| 0.025 | 10 | 50 | - | - |
| 0.004 | 30 | 105 | - | - |
| 0.001 | 50 | 185 | - | - |

**Table 1.** Response time of two samples when exposed to $H_2$ in Ar at different concentrations. Sample 1 was recorded at a low rate (2 s/point) leading to a higher signal to noise ratio but slower response time. Sample 2 was recorded at a much higher rate (0.1 s/point) leading to much faster response time. Unavailable values are marked by a dash. *: these values were recorded with a two seconds moving average.

### 3.5. Selectivity ; $H_2$ in dry or humid air.

Selectivity of a gas sensor to the analytes against other possible reactants of the medium is an important point for practical applications. Pd is known to react strongly and specifically to dihydrogen, and the good selectivity of Pd-based sensors to $H_2$ has been extensively shown in recent articles [67,68]. The carrying gas presented so far was chosen to be Ar due to its inertness to Pd, in order to investigate the intrinsic capabilities of the TAS method with comparison to previously investigated plasmonic Pd-based sensors. Nevertheless, we have verified the response of the sample to pure $O_2$, to pure synthetic dry air and to 50% relative humidity in Ar, compared to pure Ar. The results obtained on another sample are shown in fig 8.a. The effect of these three gases is very small compared to a cycle of Ar and 4% $H_2$ in Ar shown in fig 8.b. Indeed, to better see the effects, the signals were multiplied by twenty. They all show an increase in $\Delta S$ about 200 times less than that measured with 4% $H_2$. This small $\Delta S$ can be interpreted as due to the reaction of $O_2$ or of $H_2O$ on the surface of the Pd particles, leading to charge transfer between metallic Pd and the adsorbed molecules, in the same way as it has been shown previously for adsorption of oxygen and of hydrogen on gold [21,69,70].

A second important point to determine for practical application of the present system is to investigate whether it is still reactive to $H_2$ when the carrying gas is not a neutral gas but air, either dry or with water. To verify this, cycles of 4% $H_2$ in synthetic dry air and synthetic humid air (relative humidity = 50%) were undertaken and compared to cycles with 4% $H_2$ in Ar. The results are plotted in fig 8.b. Each cycle exhibits the same increase in intensity of the signal compared to the Ar cycle. The response time however is impacted by the presence of $O_2$ and of water in the carrying gas. In the case of dry air, $t_{90}$ was increased by a factor of 2 ($t_{90}$ = 50 s against 20 s in Ar). Whilst in humid air, the response time was much slower with $t_{90}$ = 250 s. Similar results which show different response times as a function of the carrying gas and of the presence of humidity have been obtained previously [68,71,72]. Yet, the detection time $t_{detect}$ in this latter case is still lower than 1s as shown in the insert of fig 8.b. This shows that although the Pd might oxidise due to the presence of $O_2$ or react with the water, its initial reactivity towards hydrogen is not inhibited, making such system efficient for security applications. A possible way for solving this issue of larger response time due to the presence of water, and improve the method for quantitative measurements in humid air, would be to protect the Pd sensor by a film porous to hydrogen but not to other molecules, especially water molecules. This has been recently achieved by using a film of metal-organic framework as a protective impermeable membrane [72].

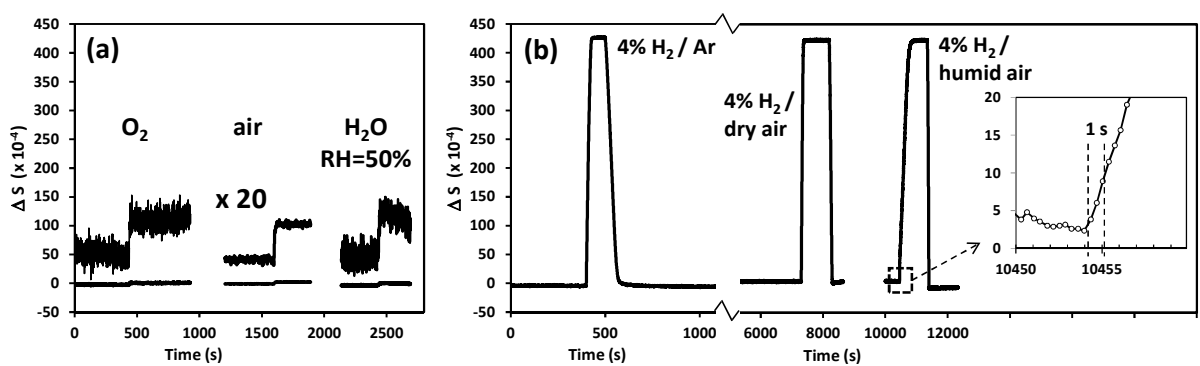

**Fig 8.** (a) Effect of other gases: switch from Ar to pure $O_2$; to pure air; to Ar with 50% relative humidity. The curves have been multiplied by a factor of 20 and vertically shifted. The noise is smaller for the 2$^{nd}$ curve, because of the use a longer integration time (1 s against 0.3 s). (b) change of the TAS signal for exposure to 4 % $H_2$ in Ar; in dry air; in air with 50% relative humidity. The time scale is different for the first curve and for the two others. The zoom shows that the detection time $t_{detect}$ is about 1 s in the latter case.

## 4. Conclusion

In this investigation, we have demonstrated the possibility of detecting as low as a few $10^{-6}$ bar, i.e. a few ppm, of $H_2$ in Ar, using an ultra-high sensitive Pd direct sensing method based on LSPR working at room temperature. This has been achieved by measuring the change of the transmission anisotropy of anisotropic nanostructured Pd film, at a single wavelength in real-time. The system presented in this paper shows a much greater sensitivity in the low concentration range of the α-phase, with respect to previous LSPR Pd sensing demonstrations, based on the determination of the spectral shift. Moreover, the resulting measurements are quantitative and permits one to determine the partial pressure of $H_2$ in a wide pressure range, from a few $10^{-6}$ bar (a few ppm) to about a few $10^{-2}$ bar. This point is an important aspect for safety consideration as it is still largely below the flammability limit of 4%. Furthermore, the time for detecting the presence of $H_2$ in the carrying gas is small, even at low partial pressure of $H_2$, and varies from 0.6 s at 4% to 50 s at 10 ppm. The simplicity of the sample elaboration is key to making this method competitive as it only consists in the evaporation at an oblique angle of a few nanometres of Pd on a glass substrate. In addition, the present optical apparatus is much less bulky compared to conventional methods as it works at a single wavelength and does not require a monochromator. Moreover, being a differential technique, it is not perturbed by ambient light. It therefore paves the way for developing light and portable miniaturised systems for easy, fast and low-cost $H_2$ sensing in a wide range of "real-world" environments. Deep investigation in real conditions is the next step to be undertaken. First results for $H_2$ in air show indeed, at least for 4% of $H_2$, the same sensitivity as $H_2$ in Ar. However, the presence of water in air increases the response time, although the detection time is still of the order of 1s. The effect of different concentrations of $H_2$ in dry air and in air with different quantities of humidity should therefore be the subject of further investigation in order to check the possible application of such Pd-based systems for developing actual $H_2$ sensors.

## Conflicts of interest

There are no conflict to declare

## Acknowledgements

The authors are grateful to Dominique Demaille for her help concerning the use of the MEB apparatus, and Sébastien Royer for technical support.

## Notes and references

# Ultra-sensitive single wavelength plasmonic hydrogen sensing with anisotropic nanostructured Pd films

## SUPPLEMENTARY MATERIAL

William L. Watkins and Yves Borensztein

Sorbonne Université, CNRS-UMR 7588, Institut des NanoSciences de Paris, F-75005, Paris, France

**S.1. Comparison between dielectric function for Pd and for PdH$_x$**

Figure S.1. shows the real and imaginary parts of the experimental dielectric function for Pd, determined by Johnson and Christy (JC) [1], and of the theoretical dielectric functions for Pd, PdH$_{0.3}$ and PdH$_{0.67}$, calculated by Silkin et al [2]. It is clear from this comparison, that the experimental $\varepsilon_{Pd}(\omega)$ is not perfectly reproduced by the calculation, although the main behaviour is correct.

When using the theoretical $\varepsilon_{Pd}(\omega)$ and $\varepsilon_{PdHx}(\omega)$ for fitting the change of the TA spectra of Figure 3.a, obtained when exposing the sample to H$_2$, there is a good agreement by using the value x=0.33 for PdH$_x$. This corresponds mainly to a decrease in intensity of both the real and the imaginary parts of $\varepsilon_{Pd}(\omega)$, as shown in Fig.S.1.b. In contrast, for PdH$_{0.67}$, the dielectric function displays a feature around 400-500 nm, which is due to interband transitions in the well-organised PdH$_{0.67}$ crystal. This is the origin of the negative feature observed in the corresponding calculated TA spectrum around 500 nm drawn in Fig.3.b. The fact that it is not observed in the experiment could be due to a random organisation of the PdH$_x$ crystal formed when the Pd NPs are exposed to pure H$_2$.

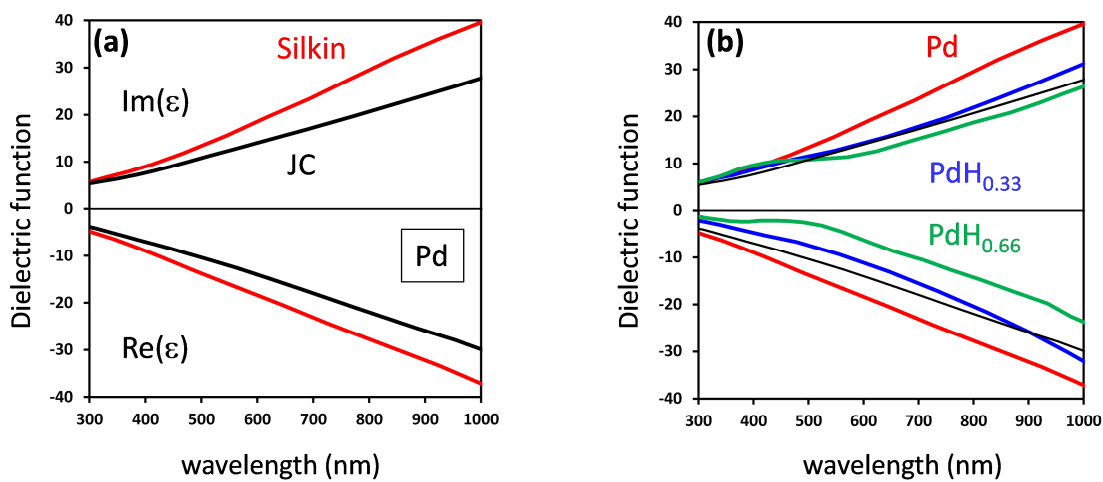

**Fig. S.1.** (a) black: experimental dielectric function of Pd from Johnson and Christy; red: dielectric function for Pd calculated by Silkin et al; (b): dielectric function calculated by Silkin et al; red: Pd; blue: PdH$_{0.33}$; green : PdH$_{0.66}$ ; thin black line : Pd from Johnson and Christy

**S.2. Modelling the optical absorption of the nanostructured anisotropic Pd film.**

Formula 1.a and 1.b are used in order to reproduce the optical absorption of the Pd film.

$$\varepsilon_x(\omega) = f\varepsilon_{Pd}(\omega) + (1-f)\varepsilon_m \qquad (S.1.a)$$

$$\varepsilon_y(\omega)^{-1} = f\varepsilon_{Pd}^{-1}(\omega) + (1-f)\varepsilon_m^{-1} \qquad (S.1.b)$$

No resonance (related to a maximum in $\varepsilon_{x,y}(\omega)$)) is obtained for parallel polarisation, whereas, for perpendicular polarisation, LSPR localised in the gaps between the interacting NPs forming the islands, are obtained for values of $\varepsilon_{Pd}(\omega)$ given by the pole of $\varepsilon_y(\omega)$, i.e. by:

$$\varepsilon_{Pd}(\omega) = -f/(1-f), \qquad (S.2)$$

with $\varepsilon_m$ taken equal to 1. This corresponds to negative values for Re($\varepsilon_{Pd}(\omega)$) which are indeed reached in the working range (Fig.S.1.a).

Fig. S.2. shows the experimental absorbance measured for perpendicular polarisation (taken from Fig.2). With a value f = 0.95, the observed maximum at 500 nm is reproduced, but the resonance is much narrower than in the experiment (blue curve). Taking into account a distribution of f, ranging from about 0.85 to 1, permits us to correctly reproduce the experimental spectrum (black curve). The distribution of f obtained in the article, where both $T_{//}$ and $T_{\perp}$ have been fitted (Fig.2), is drawn in Fig.S.3.

It is worth noticing that the Pd film could have been modelled differently. The islands formed by agglomerated NPs can also be considered as being elongated NPs with spheroidal shape, of which the LSPR is located at a wavelength which is a function of their aspect ratio. Indeed, the polarisability of such a spheroid is given by

$$\alpha_j = V \frac{\epsilon(\omega) - \epsilon_m}{\epsilon_m + L_j(\epsilon(\omega) - \epsilon_m)} \qquad (S.3)$$

and the LSPR is obtained for $\varepsilon_{Pd}(\omega) = (L-1)/L$ \qquad (S.4)

which is similar to eq S.2 where f is replaced by 1-L. Consequently, the resonance of a Pd particle with an aspect ratio equal to 0.05 is similar to the parallel slabs with f = 0.95, and is drawn in Fig. S.2 for comparison (red curve).

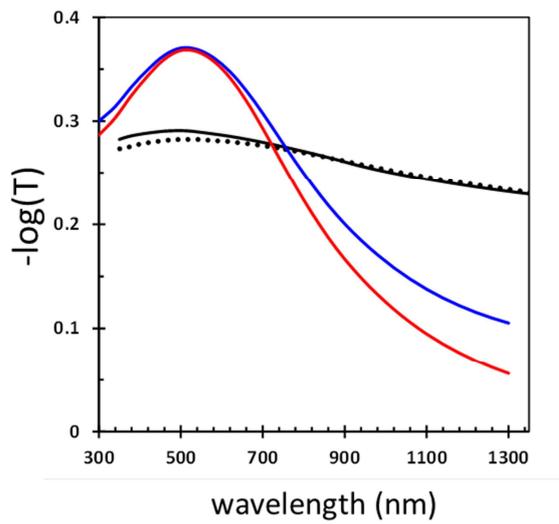

**Fig. S.2.** Absorbance. Dotted line: experimental points. Black line: fitted curve following procedure given in section S.2, with a distribution for f. Blue line: absorbance for a value of f = 0.95. Red line: absorbance for an elongated spheroid with aspect ratio = 0.05.

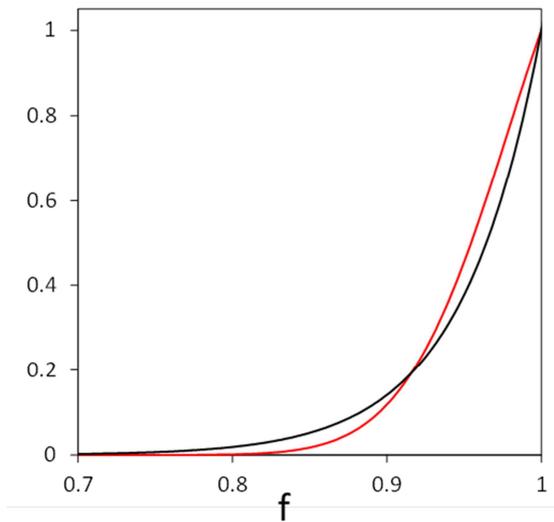

**Fig.S.3** Distribution of f values used for fitting $T_{//}$ and $T_{\perp}$ with JC dielectric function (black line), and for fitting TA with Silkin dielectric function (red line).

**S.3. Modelling the Transmission Anisotropy of the nanostructured anisotropic Pd film.**

As indicated in the main text, the TA spectrum is directly fitted by the use of the theoretical Silkin $\varepsilon_{Pd}(\omega)$, which gives a distribution of f which is only slightly different from the previous one. It is also drawn in Fig.S.3 and compared to the one obtained above. This indicates that the fitting of $T_{//}$ and $T_{\perp}$ with experimental $\varepsilon_{Pd}(\omega)$ and the fitting of TA with theoretical $\varepsilon_{Pd}(\omega)$ are in correct agreement.

### S.4. Transmission Anisotropy Spectra of different samples and response to H$_2$.

Fig. S.4.a. shows the TAS measured for three Pd samples. The optical anisotropy is expected to depend on the morphology of the samples, but an exhaustive investigation has not been done so far. The sensitivity to hydrogen is directly related to the intensity of the TAS, as it is illustrated in Fig. S.4.b for cycles of exposure to 4 % H$_2$ in Ar: the higher the optical anisotropy is, the higher the sensitivity.

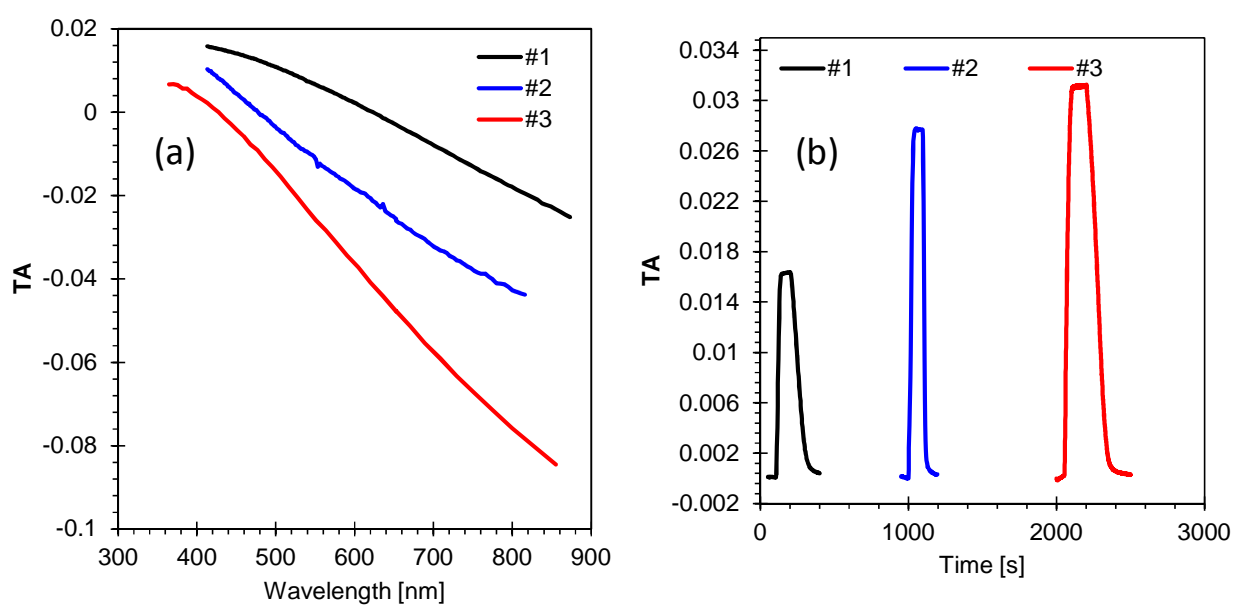

**Fig.S.4** (a): TA spectra for three Pd samples. The spectra have been shifted vertically in order to be separated on the graph. (b): Variation of the signal recorded at 885 nm, during cycles Ar / 4% H$_2$ in Ar / Ar.

### S.5. Modification of the dielectric function of Pd.

Silkin et al. calculations for the dielectric function of PdH$_x$ show a small change when x changes from 0 (pure Pd) to 0.125. In order to take into account the presence of a small amount of H in PdH$_x$, one can consider the following intermediate dielectric function, given by a linear relation, for x between 0 and 0.125:

$$\varepsilon_{PdH_x}(\omega) = \varepsilon_{Pd}(\omega) + \frac{x}{0.125}\left(\varepsilon_{PdH_{x0.125}}(\omega) - \varepsilon_{Pd}(\omega)\right) \qquad (S.5)$$

Using this dielectric function to calculate the value of the signal TA at 885 nm with the above determined distribution for f, which correctly reproduces the spectrum with pure Pd (following the procedure indicated in S.3), the values of x can be obtained for each measured value of the TA signal. The TA thus obtained are plotted in Figs 5.c and 6 of the main article.